\documentclass[traditabstract]{aa}
\usepackage{graphicx}
\usepackage{txfonts}
\usepackage{longtable}
\usepackage{xcolor}
\usepackage{natbib}
\usepackage{pdfpages}
\usepackage{verbatim}

\bibpunct{(}{)}{;}{a}{}{,}

\begin{document} 

\newcommand{\hi}{\mbox{H\,{\sc i}}}
\newcommand{\zabs}{$z_{\rm abs}$}
\newcommand{\zmin}{$z_{\rm min}$}
\newcommand{\zmax}{$z_{\rm max}$}
\newcommand{\zq}{$z_{\rm q}$}
\newcommand{\zg}{$z_{\rm g}$}
\newcommand{\kms}{km\,s$^{-1}$}
\newcommand{\cmsq}{cm$^{-2}$}
\newcommand{\degree}{\ensuremath{^\circ}}
\newcommand{\Msun}{$M_{\odot}$} 
\newcommand{\mgii}{\mbox{Mg\,{\sc ii}}} 
\newcommand{\mgiia}{\mbox{Mg\,{\sc ii}$\lambda$2796}}
\newcommand{\mgiib}{\mbox{Mg\,{\sc ii}$\lambda$2803}}
\newcommand{\mgiiab}{\mbox{Mg\,{\sc ii}$\lambda\lambda$2796,2803}}
\newcommand{\lapp}{\mbox{\raisebox{-0.3em}{$\stackrel{\textstyle <}{\sim}$}}}
\newcommand{\gapp}{\mbox{\raisebox{-0.3em}{$\stackrel{\textstyle >}{\sim}$}}}
\newcommand{\pks}{PKS\,1413$+$135}

\newcommand{\lerma}{Observatoire de Paris, LERMA, Coll\`ege de France, CNRS, PSL University, Sorbonne University, 75014, Paris --- \email{francoise.combes@obspm.fr} \label{lerma}}
\newcommand{\iucaa}{Inter-University Centre for Astronomy and Astrophysics, Post Bag 4, Ganeshkhind, Pune 411 007, India \label{iucaa}}
\newcommand{\iap}{Institut d'Astrophysique de Paris, CNRS-SU, UMR\,7095, 98bis bd Arago, 75014 Paris, France \label{iap}}
\newcommand{\chil}{French-Chilean Laboratory for Astronomy, IRL 3386, CNRS 
and U. de Chile, Casilla 36-D, Santiago, Chile \label{chil}}
\newcommand{\ioffe}{Ioffe Institute, {Polyteknicheskaya 26}, 194021 Saint-Petersburg, and HSE University, Saint-Petersburg, Russia \label{ioffe}}
\newcommand{\chalmers}{Department of Space, Earth and Environment, Chalmers University of Technology, Onsala Space Observatory, 43992 Onsala, Sweden \label{chalmers}}
\newcommand{\sarao}{South African Radio Astronomy Observatory, 2 Fir Street, Black River Park, Observatory 7925, South Africa \label{sarao}}
\newcommand{\rhodes}{Department of Physics and Electronics, Rhodes University, P.O. Box 94, Makhanda, 6140, South Africa \label{rhodes}}
\newcommand{\argl}{Argelander-Institut f\"ur Astronomie, Auf dem H\"ugel 71, D-53121 Bonn, Germany \label{argl}}
\newcommand{\nraos}{National Radio Astronomy Observatory, Socorro, NM 87801, USA \label{nraos}}
\newcommand{\nraoc}{National Radio Astronomy Observatory, 520 Edgemont Road, Charlottesville, VA 22903, USA \label{nraoc}}
\newcommand{\greenb}{Green Bank Observatory, PO Box 2, Green Bank, WV 24944, USA \label{greenb}}
\newcommand{\mpifr}{Max-Planck-Institut f\"ur Radioastronomie, Auf dem H\"ugel 69, D-53121 Bonn, Germany \label{mpifr}}
\newcommand{\rutgers}{Department of Physics and Astronomy, Rutgers, the State University of New Jersey, 136 Frelinghuysen Road, Piscataway, NJ 08854-8019, USA \label{rutgers}}
\newcommand{\uchic}{Department of Astronomy \& Astrophysics, The University of Chicago, 5640 S Ellis Ave., Chicago, IL 60637, USA \label{uchic}}
\newcommand{\labm}{Aix Marseille Univ., CNRS, CNES, LAM, Marseille, France 
\label{labm}}
\newcommand{\dura}{Institute for Computational Cosmology, Durham University, South Road, Durham, DH1 3LE, UK \label{dura}}
\newcommand{\durb}{Centre for Extragalactic Astronomy, Durham University, South Road, Durham, DH1 3LE, UK \label{durb}}
\newcommand{\tw}{ThoughtWorks Technologies India Private Limited, Yerawada, Pune 411 006, India \label{tw}}
\newcommand{\ukzn}{Astrophysics Research Centre, University of KwaZulu-Natal, Durban 4041, South Africa \label{ukzn}}
\newcommand{\ucarol}{Department of Physics and Astronomy, University of South Carolina, Columbia, SC 29208, USA \label{ucarol}}
\newcommand{\sms}{School of Mathematics, Statistics \& Computer Science, University of KwaZulu-Natal, Durban 4041, South Africa \label{sms}}
\newcommand{\idia}{The Inter-Univ. Institute for Data Intensive Astronomy (IDIA), Dep. of Astronomy, and Univ. of Cape Town, Private Bag X3, Rondebosch, 7701, South Africa, and Univ. of the Western Cape, Dep. of Physics and Astronomy, Bellville, 7535, South Africa \label{idia}}


   \title{\pks: OH and \hi\ at $z=0.247$ with MeerKAT}
   \author{
        F. Combes\inst{\ref{lerma}}
            \and
        N. Gupta\inst{\ref{iucaa}}
             \and
        S. Muller\inst{\ref{chalmers}}
           \and
        S. Balashev\inst{\ref{ioffe}} 
            \and
        P. P. Deka\inst{\ref{iucaa}}
         \and
         K. L. Emig\inst{\ref{nraoc}}
           \and
       H.-R. Kl\"ockner\inst{\ref{mpifr}}
           \and
        D. Klutse\inst{\ref{ukzn}}
           \and
       K. Knowles\inst{\ref{ukzn}, \ref{sms}}
           \and
        A. Mohapatra\inst{\ref{iucaa}}
           \and
        E. Momjian\inst{\ref{nraos}}
           \and
       P. Noterdaeme\inst{\ref{iap}, \ref{chil}}
           \and    
       P. Petitjean\inst{\ref{iap}}
         \and
       P. Salas\inst{\ref{greenb}}
            \and
        R. Srianand\inst{\ref{iucaa}}
          \and
        J. D. Wagenveld\inst{\ref{mpifr}}
           }

    \institute{\lerma \and \iucaa \and \chalmers 
    \and \ioffe \and \nraoc \and \mpifr 
    \and \ukzn \and \sms \and \nraos \and \iap \and \chil \and \greenb 
    }

   \date{Received: November 2022; accepted: December 2022}

 
  \abstract {
 The BL Lac object  \pks\ was observed by
   the Large Survey Project MeerKAT Absorption Line Survey (MALS) in the L-band, at 1139~MHz and 1293-1379~MHz, targeting the \hi\ and OH lines in absorption at $z=0.24671$. 
   The radio continuum  might come from the nucleus of the absorbing galaxy or from a background object at redshift 
   lower than 0.5, as suggested by the absence of gravitational images.
    The \hi\ absorption line is detected at a high signal-to-noise ratio, with a narrow
    central component, and with a red wing, confirming previous results. 
    The OH 1720 MHz line is clearly detected in (maser) emission, peaking at a velocity shifted by -10 to -15 \kms\ with respect to the \hi\ peak. The 1612 MHz line is lost due to radio frequency interference. The OH 1667~MHz main line is tentatively detected in absorption, but
    not the 1665~MHz line. 
    Over 30 years a high variability is observed in optical depths, due to the rapid
    changes of the line of sight caused by the superluminal motions of the radio knots.
    The \hi\ line has varied by 20\% in depth, while the OH-1720~MHz depth has varied by a factor of  $\sim$ 3. The position of the central velocity and the widths also varied.
    The absorbing galaxy is an early-type spiral (maybe S0) seen edge-on, with a prominent
    dust lane, covering the whole disk. Given the measured mass concentration and the radio
    continuum size at centimeter wavelengths (100~mas corresponding to 400~pc at $z=0.25$), the width of the absorption lines 
    from the nuclear regions are expected up to 250~\kms. The narrowness 
    of the observed lines
    ($<$ 15~\kms) suggests that the absorption comes from 
    an outer gas ring, as frequently
    observed in S0 galaxies. The millimetric lines are even narrower ($<$ 1~\kms),
    which corresponds to the continuum size 
    restricted to the core. The radio core is 
    covered by individual 1~pc
    molecular clouds, whose column density is a few 10$^{22}$ cm$^{-2}$, which is compatible with the gas screen detected in X-rays.
}
   

\maketitle
%
\section{Introduction}
\label{sec:intro}
Since its discovery \pks\ has been a mystery, raising questions regarding the classification schemes of active galactic nuclei (AGN). It is classified as a BL Lac for its highly compact variable polarized 
continuum emission, corresponding to
a relativistic jet oriented close to the line of sight \citep{Beichman1981}. \cite{Bregman1981} found only weak emission lines in the optical spectrum of the galaxy associated with the radio source, supporting the BL Lac character. They determined a redshift of $z=0.26$ for the galaxy from calcium absorption lines. However, the galaxy might not be the host of the BL Lac object.
The vast majority of BL Lac objects and radio-loud quasars are hosted by elliptical galaxies, while \pks\ appears to be hosted by an Sb-c spiral galaxy \citep{McHardy1991}. The galaxy appears almost edge-on, and  therefore the jet axis should be almost in the galaxy plane. The jet should then trigger significant activity, such as a narrow-line region, but none is detected in the galaxy. Very long baseline interferometry (VLBI) measurements have shown that \pks\ might be a compact symmetric object (CSO), a young radio source with a two-sided parsec-scale jet; more precisely, the jet structure is slightly bent, like a wide-angle tail (WAT) source \citep{Perlman1996, Perlman2002}. This means that the jet is not along the line of sight and does not benefit from beaming. However, the jet orientation might have changed with scale since the core appears beamed.
The source was also classified as a red quasar since the extinction is rather high, with a column density higher than 2$\times$10$^{22}$ cm$^{-2}$ and visual extinction A$_v >$ 30 \citep{Stocke1992}; this high extinction explains the deficit of soft X-rays from \pks. Although the radio source   coincides with the galaxy center, it was thought that the radio source might be a background source, given the absence of emission lines and strong thermal infrared flux.

\cite{Carilli1992} detected \hi\ 21  cm in absorption in the galaxy at $z=0.24671$, with a high column density ($>$ 10$^{22}$ cm$^{-2}$), supporting the high extinction found by \cite{Stocke1992}, and confirming the classification of the galaxy as a late-type spiral. Molecular clouds were also observed in absorption by \cite{Wiklind1994}, supporting a high gas column-density. Monitoring over two years revealed high variation 
in the molecular absorption opacity spectrum, attributed to the superluminal character of the BL Lac. This high gas column density exacerbates the problem of the absence of thermal emission expected from the gas heated by an internal active nucleus. 
An alternative interpretation of the variability of BL Lac objects is that the background AGN suffers microlensing by the stars of a foreground galaxy \citep{Ostriker1990}. For better efficiency of stellar microlensing, while avoiding macrolensing, the AGN should be close to the foreground galaxy.
The Very Long Baseline Array (VLBA) data from \cite{Perlman2002} provide a high resolution map of the HI absorption, which is mainly in front of the eastern mini-jet 25~mas from the radio core. The molecular gas absorbs in front of the core, and may be also the OH line.
The core happens to be offset by
13~mas (=50~pc) from the gravity center of the spiral galaxy. This offset might favor a background location for the AGN, although \cite{Perlman2002} favor a patchy absorption from the spiral galaxy. If the AGN is behind the spiral, it 
 must be at z $<$0.3 to avoid arcs and lensing images, which are not observed, and thus close to the spiral ($z=0.25$) we should see its galaxy host in HST images.

Recently, \cite{Vedantham2017a, Vedantham2017b} discovered symmetric achromatic variability (SAV) in \pks\ in the forms of  a 1yr long U-shaped perturbation of the light curve, with a symmetric feature 5 yr later.
This nicely corresponds to the signature of gravitational milli-lensing when knots in the relativistic jet cross caustics created by lenses of mass between 10$^3$ and 10$^6$ M$_\odot$ within the intervening galaxy. Combining all VLBI monitoring at several wavelengths, \cite{Readhead2021} conclude that \pks\ is a background BL Lac quasar, with radio jets aligned along the line of sight and located at
0.25 $<$ z $<$ 0.5. The intervening galaxy is a  nearly edge-on Seyfert 2 spiral, providing the milli-lensing episodes responsible in part for the variability, the other part being intrinsic to the BL Lac. They show that the jet appears quasi-perpendicular to the galaxy major axis, and has an extension of 110~mas.
Since the optical spectrum from \cite{Vedantham2017a} should reveal signatures from the background galaxy, which are not detected, the localization of the radio continuum source 
 is not yet settled.  It is possible that the BL Lac 
 is not hosted by a massive elliptical, but by a galaxy half as luminous as the
 foreground Seyfert;  thanks to  the high dust obscuration observed
 (by a factor of $\sim$10 in the near-infrared) and the luminosity dimming at z=0.5, the expected flux ratio between the two galaxies could be as high as 100.

The size of the radio source at centimeter wavelengths is 0.11'' $\sim$ 400pc \citep{Perlman1996}. This 
broadens the absorption spectrum because the gas actually involved in the absorption traces the velocity gradient over that length in the disk for \hi\ 21cm and OH 18cm. The millimeter lines are much narrower  since the continuum is only significant in the core \citep{Wiklind1994}.

\cite{Kanekar2002} detect the main OH line at 1667 MHz, while \cite{Darling2004} detects the conjugate satellite lines (1612 MHz in absorption and 1720 MHz in emission), but none of the main OH lines. 
The 1720 and 1612 MHz line rest frequencies have different  dependences on the fine structure constant $\alpha$ and the proton-electron mass ratio $\mu$, and they have been used 
to constrain the variation of these fundamental constants \citep{Kanekar2002,Kanekar2004, Kanekar2018, Darling2004}. Because the lines are conjugate, and their optical depths are perfectly symmetric, they are a privileged tool, eliminating the possible biases 
that the lines are not coming from the same gas along the line of sight. In the present work we report new \hi\ and OH observations of \pks\ carried out as part of the MeerKAT Absorption Line Survey \citep[MALS; ][]{Gupta17mals}. We quantify the time variability of the various components, and through comparison with previous millimeter-wave observations we  draw some conclusions on the nature and location of the absorptions.

This paper is structured as follows. Section \ref{sec:obs} presents the details of the observations and data analysis with {\tt ARTIP}. The results in terms of optical depth and column density are quantified in Section \ref{sec:res}, as well as the time variability.
 We discuss in
Sect. \ref{sec:kinmod} the origin of the various absorption components in comparison with the dense molecular gas absorptions. Section \ref{conclu} summarizes our conclusions.
Throughout the  paper the velocity scale is defined with respect to $z=0.24671$ (barycentric reference frame), which corresponds to the main atomic \citep{Carilli1992} and
molecular (e.g., CO, HCO$^+$, HCN, and OH) absorption components detected at millimeter  wavelengths \citep[][]{Wiklind1994} or centimeter wavelengths \citep{Kanekar2002, Darling2004}.
To compute distances we adopt a flat $\Lambda$CDM cosmology,
with $\Omega_m$=0.29, $\Omega_\Lambda$=0.71, and the Hubble constant
H$_0$ = 70~\kms Mpc$^{-1}$. At the distance of the \pks\ absorber, 1 arcsec corresponds to 3.9\,kpc in physical units.

\section{Observations and data analysis}
\label{sec:obs}

The field centered at \pks\ was observed as part of MALS on October 16, 2020, using 59 antennas of the MeerKAT-64  array and the  32K mode of the SKA Reconfigurable Application Board (SKARAB) correlator. The total bandwidth of 856 MHz centered at 1283.9869 MHz was split into 32,768 frequency channels. The resultant frequency resolution is 26.123 kHz, or 6.1\,\kms, at the center of the band.  The correlator dump time was 8\,seconds. The objects  3C286 and PKS1939–638 were observed for flux density scale, delay, and bandpass calibrations.  The compact radio source J1347+1217 was observed for complex gain calibration.  The total on-source time on \pks\ was 56\,mins. 
The data were processed using the latest version of the Automated Radio Telescope Imaging Pipeline {\tt ARTIP} based on CASA 6.3.0 \citep[][]{Casa22}. The details of {\tt ARTIP} and the  calibration steps are provided in \citet[][]{Gupta2020}.  

The spectral line processing of the  MALS datasets through {\tt ARTIP} partitions the wideband into 16 spectral windows (SPWs) labeled   SPW-0 to SPW-15 \citep[for details, see][]{Gupta2020}.  
In short, the measurement sets for each SPW are processed for continuum imaging with self-calibration.  We performed three rounds of phase-only with a solution interval of 1\,min,  and one round of amplitude-and-phase self-calibration with a solution interval of 5\,min. These self-calibration gain solutions were applied to the visibilities prior to continuum subtraction and cube imaging.
The \hi\ 21cm and the four OH lines for $z=0.24671$ would fall in SPWs 4, 8, and 9.
Here, we mainly focus on the stokes\,$I$ radio continuum and spectral line properties of the target obtained using the {\tt robust = 0} weighting of the visibilities for these SPWs, as implemented in CASA. 
The frequency ranges covered by these SPWs and the corresponding spectral rms are summarized in Table~\ref{tab:spwsum}.  In the table we also provide the synthesized beams and continuum flux densities for the continuum images generated using the visibilities from these SPWs. 

\begin{table*}
\caption{Properties of PKS1413+135 observations for relevant SPWs.}
\vspace{-0.4cm}
\begin{center}
\begin{tabular}{ccccc}
\hline
\hline
{\large \strut} SPW-Id. & Frequency range & Spectral rms    &  Synthesized beam & Total flux density$^a$ \\
                        &     (MHz)       &(mJy\,beam$^{-1}$) &                  &     (Jy)       \\         
\hline
SPW-4$^b$    & 1083.4 - 1143.6  &  1.0    &  $16\farcs7\times8\farcs1$, $-21.8^\circ$   &  1.3635$\pm$0.0032   \\
SPW-8$^c$    & 1297.4 - 1357.6  &  0.9    &  $14\farcs0\times6\farcs8$, $-22.7^\circ$   &  1.1682$\pm$0.0067   \\
SPW-9$^d$    & 1350.9 - 1411.1  &  0.9    &  $13\farcs5\times6\farcs5$, $-22.8^\circ$    & 1.1335$\pm$0.0056     \\
\hline
\end{tabular}
\tablefoot{ 
\tablefoottext{a}{Obtained using a single-Gaussian component fitted to the radio emission.}  
\tablefoottext{b}{Covers the  \hi\ 21cm line.} 
\tablefoottext{c}{Covers the  OH 1612, 1665, and 1667\,MHz lines.}   
\tablefoottext{d}{Covers the  OH 1720\,MHz line.}   
        }
\label{tab:spwsum}
\end{center}
\end{table*}

\begin{figure}
\begin{center}
\includegraphics[width=0.51\textwidth,angle=0]{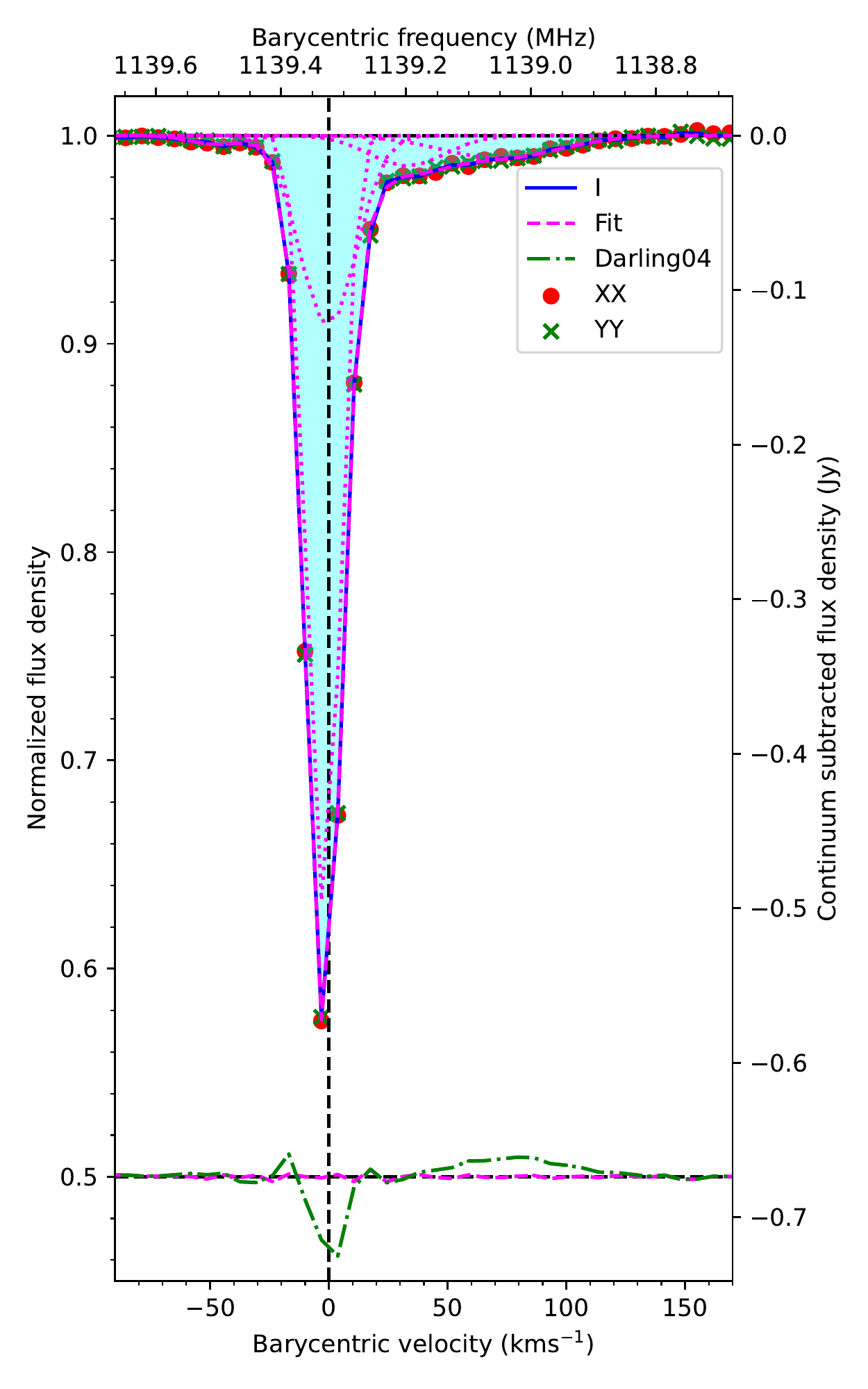}
\vskip+0.0cm
\caption{
        MeerKAT stokes $I$ \hi\ 21cm spectrum of \pks.  The individual Gaussian components fitted to the absorption profile (dotted lines), the total fit (dashed line), and the residuals (dot-dashed lines) are also shown.
        For comparison, we also show the residuals with respect to the GBT spectrum reproduced using the Gaussian components provided in \cite{Darling2004} (dot-dashed green line).    
        The individual XX and YY MeerKAT spectra are shown as points and crosses, respectively.    
} 
\label{fig:pks1413-HI} 
\end{center}
\end{figure} 

\section{Results}
\label{sec:res}

\subsection{\hi\ absorption}

\begin{figure}
\centering
\includegraphics[width=0.48\textwidth,angle=0]{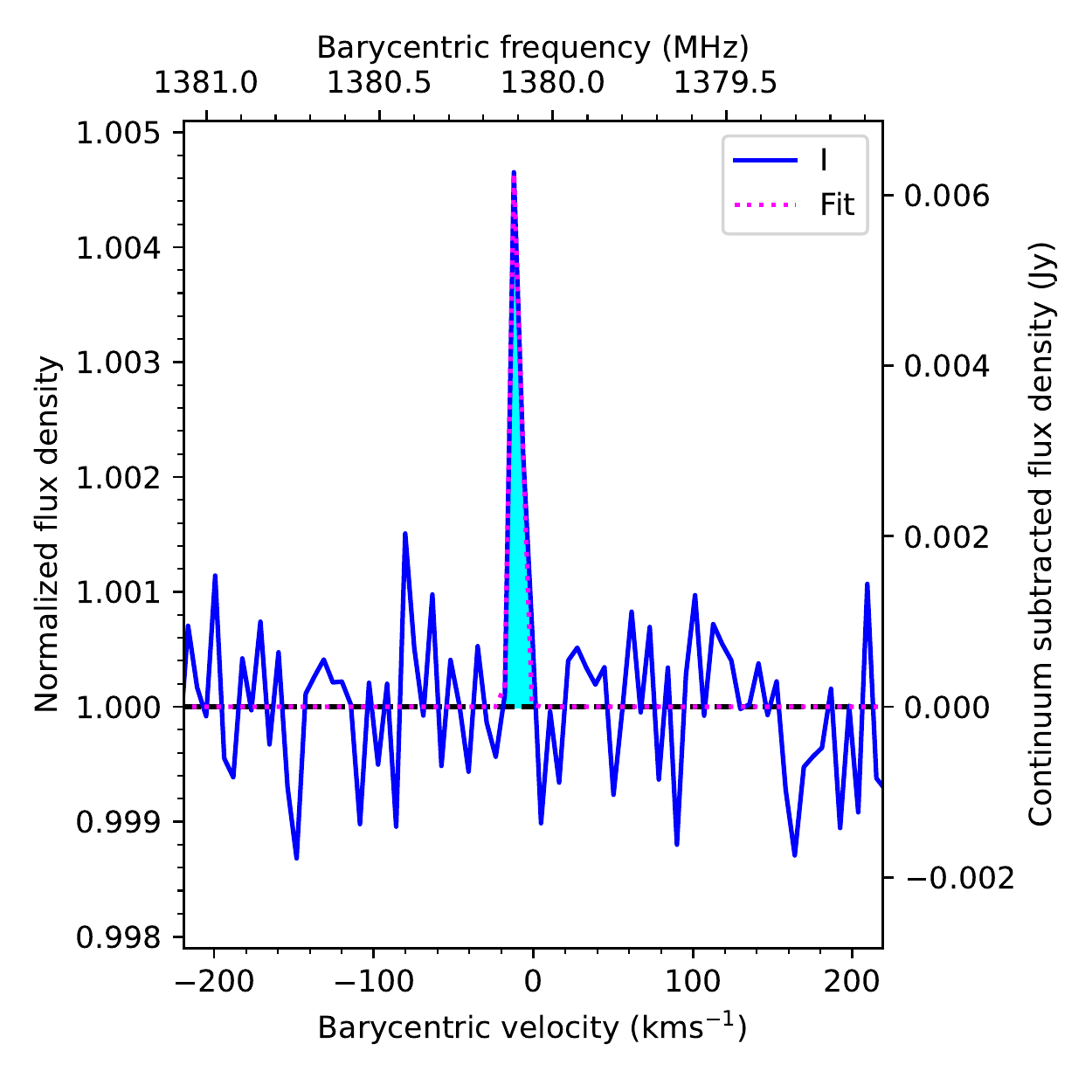}
\vskip+0.0cm
\caption{
MeerKAT Stokes-$I$ spectrum (blue) of OH 1720 MHz line at $z=0.24671$ toward \pks.  The single-Gaussian component (peak = 0.0076$\pm$0.0031\,Jy; shift = -10.37$\pm$0.99\,\kms; sigma = 2.8$\pm$1\,\kms) fitted to the emission line is shown as a dashed line. The emission corresponds to a weakly masing line, while the 1612 MHz emission is seen in absorption \citep{Darling2004}. The RFI at this frequency prevented MeerKAT from observing the emission. 
}
\label{fig:pks1413-1720}
\end{figure} 

Figure~\ref{fig:pks1413-HI} displays MeerKAT \hi\ 21cm absorption in front of
\pks\, at $z=0.24671$. Given the asymmetry of the line, the best fit providing a satisfactory model of the total spectrum required  five Gaussian components. The characteristics of all the components are listed in Table~\ref{tab:gaussfit}. These components may not be physical, but are just practical to completely reproduce the global profile.
There is a wide component in the red wing, which was also   seen by 
\cite{Darling2004}, although it was thought at that time that it might  come from the radio frequency interference (RFI).
The global integrated \hi\ 21cm optical depth is $\int\tau$dv$ = 10.86\pm 0.32$\,\kms.
Assuming a homogeneous and optically thin gas, at spin temperature 
$T_{\rm s}$= 100K, with a covering factor $f_c^{\tiny \hi}$ = 1, and using the equation
\begin{equation}
        N{(\hi)}=1.823\times10^{18}~{T_{\rm s}\over f_{\rm c}^{\tiny \hi}}\int~\tau(v)~{\rm d}v~{\rm cm^{-2}}
\label{eq21cm}
\end{equation}
we obtain the \hi\ column density $N$(\hi) = $(1.98 \pm 0.06)\times10^{21}$ \cmsq.

\begin{table}
\caption{Multiple Gaussian fits to the \hi\ 21cm absorption.}
\vspace{-0.4cm}
\begin{center}
\begin{tabular}{cccc}
\hline
\hline
{\large \strut}     ID    &    Center     &  $\sigma$            &   $\tau_p$       \\
                           &   (\kms)      &   (\kms)             &                  \\
\hline
    1      &  -47.1 $\pm$ 2.1  &  11.1 $\pm$  2.2   &  0.00434 $\pm$  0.00057\\
    2      &  32.9  $\pm$ 4.0  &  15.9 $\pm$  5.2   &  0.0147  $\pm$  0.0045 \\
    3      &  72.1  $\pm$ 9.2  &  24.9 $\pm$  4.8   &  0.0108  $\pm$  0.0017 \\
    4      &  -2.19 $\pm$ 0.024&  6.35 $\pm$ 0.064  &  0.463   $\pm$  0.015  \\
    5      &  -0.322$\pm$ 0.86 &  11.7 $\pm$ 0.89   &  0.0963  $\pm$  0.013  \\
\hline
\end{tabular}
\label{tab:gaussfit}
\end{center}
\end{table}

\subsection{OH main and satellite lines}
\begin{figure*}
\includegraphics[width=0.95\textwidth,angle=0]{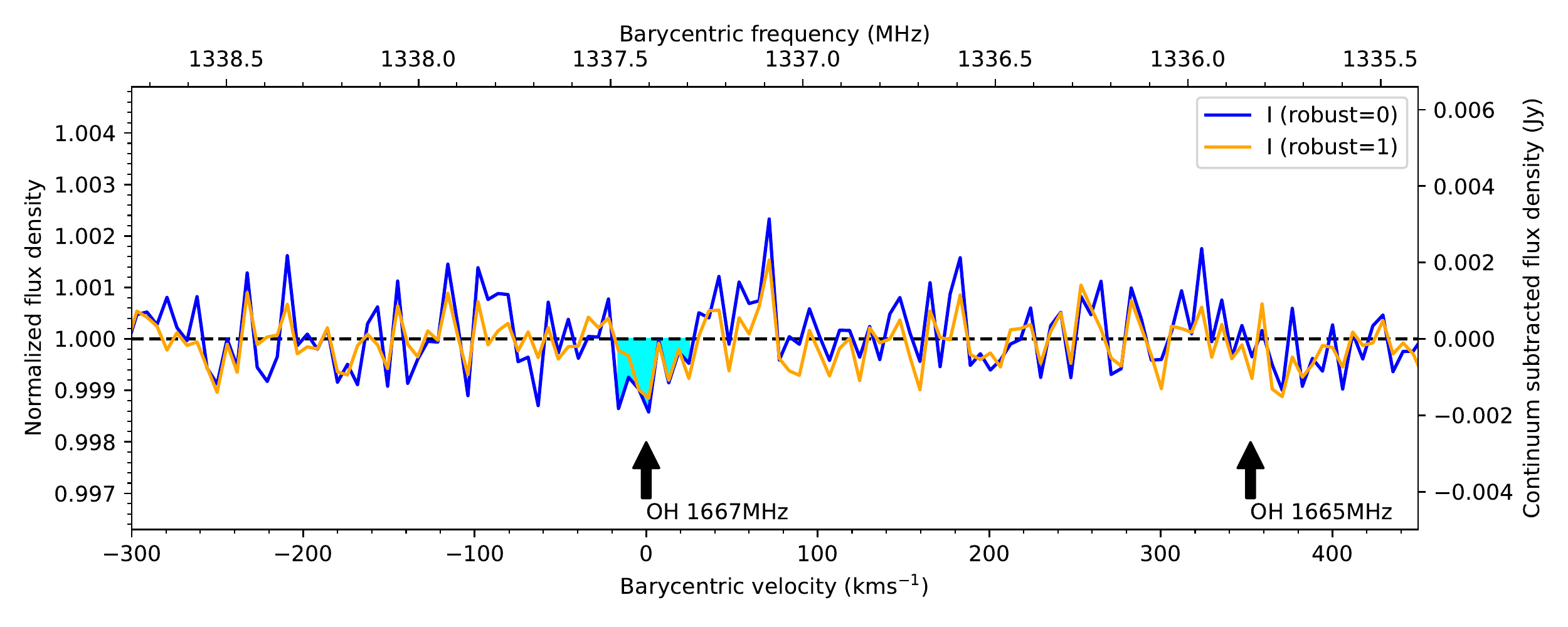}
\vskip+0.0cm  
\caption{
MeerKAT  Stokes-$I$ spectrum (blue for robust=0 and orange for robust=1) of the OH main lines absorption at $z=0.24671$ toward \pks. The spectral rms noise is 0.9\,mJy\,beam$^{-1}$\,channel$^{-1}$. Vertical arrows indicate the expected positions of the 1665 and 1667 MHz lines. There is a tentative dip in the flux at the location of 1667\,MHz line corresponding to $\int\tau$dv=0.03$\pm$0.01\,\kms. 
} 
\label{fig:pks1413-OH}
\end{figure*}

Figure \ref{fig:pks1413-1720} shows that the 1720~MHz satellite line is well detected with
MeerKAT, while Fig. \ref{fig:pks1413-OH} displays the spectrum of the OH main lines. There is only a hint of a detection of the 1667~MHz line, which was already detected by 
\cite{Kanekar2002}. The 1665~MHz line is not detected.
In local thermodynamic equilibrium (LTE), the relative strengths of the  18cm lines are expected to follow the ratio 1612:1665:1667:1720\,MHz = 1:5:9:1. Since we tentatively detect the 1667~MHz line, there is no surprise that there is only an upper limit for the 1665~MHz line.

We detect the 1720~MHz line in emission (see Fig. \ref{fig:pks1413-1720}), while the expected absorption at 1612~MHz is undetected due to RFI. The two satellite lines have been clearly detected by \cite{Darling2004} and \cite{Kanekar2018} to be exactly symmetric, with their sum of negative and positive integrated optical depths cancelling out.  The OH satellite lines were also observed to exhibit conjugate behavior in PKS1830-211 \citep{Combes2021}.

\begin{table}
\caption{Multiple epochs of HI and OH absorptions at $z=0.24671$}
\vspace{-0.4cm}
\begin{center}
\begin{tabular}{cccc}
\hline
\hline
{\large \strut} Epoch (HI) &    S(Jy)  &  $\tau_{HI}$   &   Ref      \\
         1992    & 1.25$\pm$0.15  &   0.34 $\pm$ 0.04 &   (1)    \\ 
         2003   & 1.37$\pm$0.02   &   0.68 $\pm$ 0.03 &   (2)    \\ 
         2020   & 1.363$\pm$0.003  &  0.463 $\pm$ 0.015 &   (3)    \\
\hline
{\large \strut} Epoch (OH) &    S(Jy)  &  $\tau_{1720}$      &   Ref      \\
         2003   & 1.085$\pm$0.002 & 0.0097 $\pm$ 0.0017   & (2)    \\ 
         2003   &               &  0.014 $\pm$ 0.002   &   (4)    \\ 
       2010-12   &          &   0.0071 $\pm$ 0.0003   &   (5)  \\ 
         2020   & 1.133$\pm$0.006  &  0.0047 $\pm$ 0.0018 &   (3)    \\
\hline
\end{tabular}
\tablefoot{ 
\tablefoottext{a}{ Ref (1): \cite{Carilli1992} VLA with resolution 1.28~\kms. 
Ref (2): \cite{Darling2004} GBT with 0.8~\kms. Ref (3): this work.
Ref (4): \cite{Kanekar2004} WSRT with 1.1~\kms.
Ref (5): \cite{Kanekar2018} Arecibo with 0.18~\kms.}
        }
\label{tab:HIOHvar}
\end{center}
\end{table}

\subsection{Time variability}
\label{sec:var}

\begin{figure}
\includegraphics[width=0.48\textwidth,angle=0]{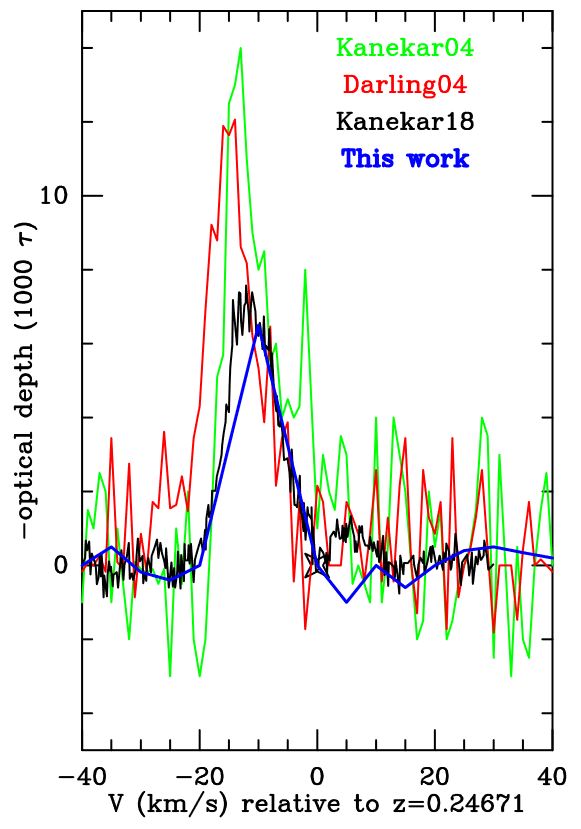}  
\vskip+0.0cm  
\caption{  
MeerKAT  OH 1720~MHz spectrum  taken in 2020 (blue), compared to the spectrum  observed in 2003 by \cite{Kanekar2004} with WSRT (green), the spectrum observed in 2003 by \cite{Darling2004} with GBT (red),
 and the spectrum observed in 2010--2012 by \cite{Kanekar2018} with Arecibo (black).
All spectra have been normalized to their observed continuum.
} 
\label{fig:comp-OH}   
\end{figure}

The radio continuum emission from the background blazar is found to  vary in time, sometimes with large factors, likely due to milli-lensing \citep{Peirson2022}.
 However, the optical depth of the \hi\ and OH absorption lines  vary even more (see Table \ref{tab:HIOHvar} and Fig. \ref{fig:comp-OH}).

Figure~\ref{fig:pks1413-HI} shows that the difference in the \hi\ optical depth between \cite{Darling2004} and the present work is detected at 6$\sigma$, while the  
peak optical depth in 2003 was twice that of 1992 (see Table \ref{tab:HIOHvar}). The bulk of the \hi\ absorption does not come from the core of the radio source, but from one knot in the radio jet, the eastern mini-lobe at 25~mas (=100~pc)  from the core, as shown with the VLBA by \cite{Perlman2002}. Since the jet has superluminal motions, it is expected that the line of sight explored is varying relatively quickly, especially in a clumpy medium \citep{Wiklind1997}. The continuum source is not extended more than a few tens of milliarcseconds, and the absorbing medium has the size of an interstellar cloud.

We compared all available OH 1720~MHz spectra from 2003 with the present MeerKAT spectra. Figure \ref{fig:comp-OH} shows that     the peak optical depth is varying, as are  the central velocity and the width of the spectral features. The absorption peaks between 5\% and 14\% of the continuum, with a central velocity from -10 to -15~\kms\, with respect to the CO and \hi\ velocities. The FWHM varies between 7 and 11~\kms. 
Although observations at a smaller number of epochs are available for the \hi\ absorption, it appears that it is slightly less variable, as expected for a less clumpy component. As already noted in \cite{Combes2021} for PKS1830-211, the atomic component is more diffuse, more extended both in radius and in height in the galaxy plane.

Additional sources of variation might be due to the milli-lensing events, 
observed through the symmetric achromatic variability events monitored by \cite{Vedantham2017a, Vedantham2017b}. These events usually last more than one year, and they occurred clearly in 1993--1995, 2009--2010, 2014--2015 \citep{Peirson2022}. They also affected some absorption measurements in 2010 \citep{Kanekar2018}, and also in the molecular component \citep{Wiklind1994, Wiklind1997}.

\section{Kinematic interpretation of the spectra}
\label{sec:kinmod}

\pks\, has been considered for a long time as a puzzling source. It shows two sets
of characteristics that appear contradictory. On one side it exhibits a BL Lac character
with a radio jet relativistically boosted toward the observer, with apparent transverse velocity up to v$\sim$ 1.5c measured through proper motions with VLBA 
\citep[e.g.,][]{Lister2016}. This would mean that
the jet orientation is close to the line of sight; however, it also has a counterjet, which suggests a re-orientation somewhere along the jet. Paradoxically, it has also been classified as a CSO, which  are typically small and young objects
that have jet axes not aligned close to the line of sight. CSOs are often thought to have re-starting jet activity, and \pks\    in 2019 revealed violent variations (more than an order of magnitude) in $\gamma$-rays \citep{Principe2021}. It is the first CSO detected at
very high energies, larger than 100 GeV \citep{Blanch2022}. \cite{Gan2022} show that the source becomes harder in frequency when brighter, and that the recent flare of PKS1413+135 might be linked to its re-started jet activity.

Powerful radio jets are not generally associated 
with spiral galaxies, and 
in the case of \pks\, the radio source core corresponds to the nucleus of an
edge-on early-type spiral galaxy \citep{McHardy1994}. Many authors
\citep[e.g.,][]{McHardy1991,Perlman2002} have questioned its association
with this galaxy since no infrared activity is observed,
contrary to what is expected 
for the host of a powerful AGN that has a radio jet aligned along its plane. The true galaxy host could be a background galaxy, hard to detect since hidden by the foreground edge-on galaxy. The latter could 
have some lensing effect on the BL Lac, but since no lensed images are seen,
its redshift
{\bf may be} limited to the range 0.2467 $<$ z $<$ 0.5 \citep{Readhead2021}. Only milli-lensing events are possible, and would correspond to the SAV events
observed by \cite{Vedantham2017a}. 

The various absorption lines coming from the foreground edge-on galaxy and observed in front of the background radio source provide precious information. This data 
could bring new clues to the nature of this puzzling source.

\subsection{Galaxy model}

The \hi\ spectrum is remarkable for its main narrow central component at zero velocity,
and its redshifted wing, extending up to 100~\kms. It is also striking that it is not seen toward the radio core, but mainly from a counterjet knot, 25ms (=100~pc) northeast from the core \citep{Perlman2002}, on the minor axis of the galaxy.
All the other absorption lines come from the core region, where the X-ray absorption
indicates a large column density N(H) $\sim$ 6 $\times$ 10$^{22}$ cm$^{-2}$ \citep{Stocke1992}.

The systemic velocity was set such that the main narrow \hi\ absorption
is centered at zero velocity. All lines peak at this velocity, except the OH 18cm satellite lines. This might   also be the systemic velocity of the foreground
galaxy since
most of the absorption is coming from the minor axis of the rotating spiral.
On the contrary, the two OH satellite lines appear at V=-10 or -15~\kms, according to the observing epoch (see Fig. \ref{fig:comp-OH}).
The fact that the \hi\ absorption does not come from the core (almost coincident 
 with the foreground galaxy nucleus)
 is not surprising since there is a well-known   \hi\ depletion in spiral galaxy
 centers \citep[e.g.,][]{Walter2008}.

\begin{figure}
\includegraphics[width=0.48\textwidth,angle=0]{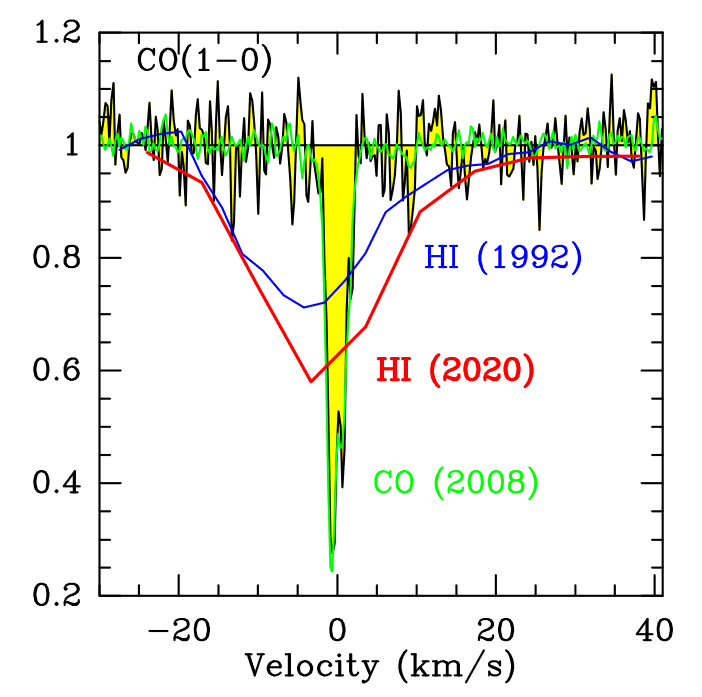}  
\vskip+0.0cm  
\caption{Comparison between the CO(1-0) and \hi\ lines. All spectra are normalized to their respective continuum. The blue curve is the \hi\ absorption, observed in 1992 with the VLA by \cite{Carilli1992} with channels 2.57~\kms, and the red curve is from this work (\hi\ spectrum obtained in 2020 with 6~\kms\ channels). Superposed on the CO(1-0) spectrum observed in 1996  (black) from \cite{Wiklind1997} is   a CO(1-0) spectrum (green) obtained with the Institut de Radioastronomie Millim\'etrique (IRAM) interferometer in 2008, showing no variation.} 
\label{fig:comp-HI-CO}   
\end{figure} 

The edge-on spiral galaxy appears to be of very early type, perhaps a lenticular. 
This is supported by the large and massive bulge, as photometrically fitted
by \cite{McHardy1994}. With the bulge alone the effective radius is best fit as 11~kpc, but when the bulge and disk are fitted together the effective bulge radius is found to be
9.7~kpc. The disk is quite thick,
with characteristic height-to-size ratio h/R = 0.24. The mass concentration 
is high, and a Sersic index of n=4 is used. With this  mass concentration,
for a galaxy of total mass $\sim$ 3 $\times$ 10$^{10}$ M$_\odot$
the velocity width at the nucleus is expected to be on the order of 300~\kms.

Since all absorption lines are quite narrow, with a FWHM lower than 
15~\kms, except for the weak red wing of the \hi\ spectrum,
it is likely that the absorbing gas is not located in the nuclear region, but 
is orbiting in an outer gaseous ring. These gaseous rings are frequently observed
in lenticular galaxies \citep[e.g.,][]{vanDriel1991}. The ring likely corresponds 
to the dust lane observed  crossing the disk in the HST image of \cite{McHardy1994}.
Because the dust lane covers the whole disk extent, it must be an outer ring.

\subsection{Nature of the absorbing gas}
 The \hi\ 21cm and OH 18cm  lines are narrow, but the millimetric lines
 are even narrower. As reported by \cite{Wiklind1997}, the CO(1-0) line
 is composed of two main components, of FWHM = 0.6-1.9~\kms, the deepest
 absorption being at velocity V=-0.5~\kms\ and the other  at +0.6~\kms. 
 According to the epoch observed, a third weaker line
 might appear wider (FWHM 0.6-7~\kms) in the red wing located at V=1-2~\kms. The position,
 intensity, and widths of these components   varying with time, by factors  of $\sim$2
 over two years, due to the change in line of sight toward the continuum core, with the knots moving
 at superluminal speeds.

\begin{figure}
\includegraphics[width=0.48\textwidth,angle=0]{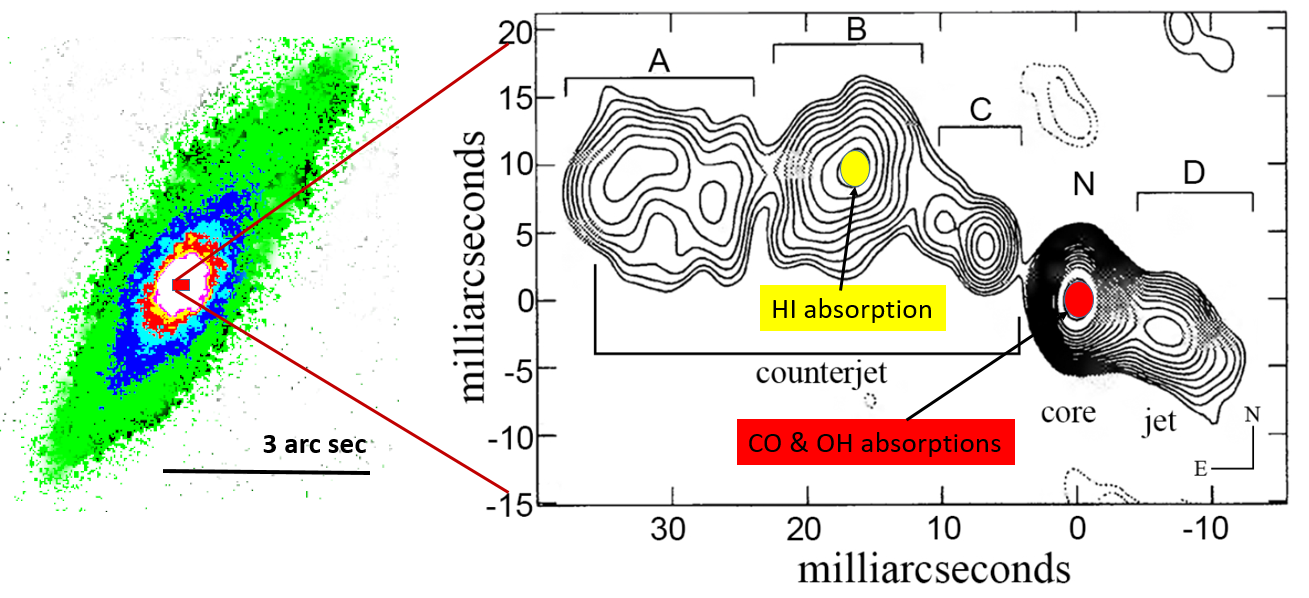}  
\vskip+0.0cm  
\caption{Schematic view of the atomic and molecular absorptions. The HST legacy archive image of the absorbing galaxy in H-band (F160W) is plotted on the left, and the 5 GHz VLBI image from \cite{Perlman1996} showing the jet (west) and counterjet (east) is on the right. The contours are spaced by a factor of $\sqrt 2$ in brightness. The component N is the nucleus or radio core. The counterjet components are indicated by A, B, and C.  The HI absorption is centered in B, and the molecular components in N. The jet structure is aligned on the minor axis of the galaxy.} 
\label{fig:pks1413-schema}   
\end{figure}

 Figure \ref{fig:comp-HI-CO} highlights the difference in widths between 
 the centimeter and millimeter absorption lines. This is easily interpreted due to
 the difference in extent of the background continuum at these two frequencies.
  As shown in \cite{Perlman1996, Perlman2002}, the radio source contains a jet and counterjet, of size 100~mas $\sim$ 400~pc,
  but only at centimeter wavelengths. Already at 43 GHz the source is reduced to a point-like
  core, with the beam size ($<$ 1~mas). The millimeter pencil beam of the continuum
  may encounter only one cloud, of size 1~pc, corresponding to the observed width,
  according to the size--linewidth relation \citep{Solomon1987}. 
  The medium is certainly clumpy,
  and several components may occur on the edge-on gaseous ring, for 
  a kiloparsec width of the ring. For \hi\ and OH absorption lines the continuum
  covers a broader region of the ring, explaining the observed wider absorptions.
  A schematic view of the various components is sketched in Fig. \ref{fig:pks1413-schema}.
  
  Figure \ref{fig:comp-HI-CO} also shows that the CO(1-0) spectrum has not varied between 1996 and 2008. This might indicate that the core component, toward which the millimeter absorptions are detected, is not actually moving, and that the superluminal
  motions only affect the jet. At centimeter wavelengths, the absorptions occur toward the more extended  continuum emission, including the jets, for both OH and \hi\,, and this
  explains their variations. Although the \hi\ and OH absorptions are not centered at the same point, their extensions overlap, and that is why their variations are somewhat correlated:  their optical depths both decreased between 2003 and 2020.

\section{Conclusions}
\label{conclu}
   We report results from the MALS project on the MeerKAT array toward the
   BL Lac source \pks. A strong \hi\ absorption is detected, decomposed into five
   Gaussian components, with high signal-to-noise ratio (S/N$\sim$ 30 in channels of 6~\kms\,).
   The main OH component at 1667~MHz is tentatively detected, but not the 1665~MHz component. The 1720~MHz satellite line is clearly detected, while radio frequency interference (RFI) prevented us from detecting the expected absorption line at 1612~MHz.
   The 1720~MHz satellite line is observed in stimulated (maser) emission, due to radiative pumping through far-infrared radiation. Previous observations \citep{Darling2004, Kanekar2018} have shown that the two satellite lines are perfectly conjugate, with the 1612~MHz 
   line in absorption.
   The absorption lines have been observed   since 1992. The \hi\ component
   reveals a 30\% variation in its depth in 2020 with respect to 2003 (see Table
   \ref{tab:HIOHvar}), but only slightly in velocity position or width. The
   1720~MHz OH satellite line shows more variation: a factor of 3 in optical depth,
    in velocity position, and in width (see Fig. \ref{fig:comp-OH}).
   These variations can easily be explained by the change in the line of sight,
   related to the superluminal motion of the background radio continuum knots.
   
   Due to discovered SAV in \pks,
   attributed to milli-lensing events, it is possible that 
   the continuum source belongs
   to a background AGN, at redshift lower than 0.5 \citep{Peirson2022}.
   The size of the jet and counterjet
   covers 400~pc at the distance of the foreground absorbing 
   galaxy in the rest frame 18 and 21cm wavelengths.
   The very narrow (FWHM $<$ 15~\kms) OH and \hi\ absorption lines suggest
   that the absorbing interstellar medium in the foreground edge-on galaxy 
   is located in an outer gas ring. The foreground galaxy is an early-type spiral,
   likely a lenticular given its very massive bulge. The width
   sampled by the gas in the 400~pc nuclear region is expected to be much larger
   (250~\kms). In the millimeter domain, the continuum size is much
   smaller ($<$ 4~pc), restricted to its core. Along this pencil beam
   the line of sight may encounter individual parsec-scale clouds, accounting for the
   extremely narrow (FWHM$\sim$ 1~\kms) absorption in CO, HCO$^+$, or HCN
   \citep{Wiklind1997}. The typical surface density of molecular clouds (a few 10$^{22}$ cm$^{-2}$) corresponds to the optical depth found
   in X-ray data \citep{Stocke1992}.

\begin{acknowledgements}
The MeerKAT telescope is operated by the South African Radio Astronomy Observatory, which is a facility of the National Research Foundation, an agency of the Department of Science and Innovation. KM acknowledges support from the National Research Foundation of South Africa.
The MeerKAT data were processed using the MALS computing facility at IUCAA (https://mals.iucaa.in/releases).
The Common Astronomy Software Applications (CASA) package is developed by an international consortium of scientists based at the National Radio Astronomical Observatory (NRAO), the European Southern Observatory (ESO), the National Astronomical Observatory of Japan (NAOJ), the Academia Sinica Institute of Astronomy 
and Astrophysics (ASIAA), the CSIRO division for Astronomy and Space Science (CASS), and the Netherlands Institute for Radio Astronomy (ASTRON) under the guidance of NRAO.
The National Radio Astronomy Observatory is a facility of the National Science Foundation operated under cooperative agreement by Associated Universities, Inc. We made use of the
NASA/IPAC Extragalactic Database (NED).

\end{acknowledgements}

\bibliographystyle{aa}
\bibliography{PKS1413.bib}

\end{document}